\documentstyle[newarcrc,fleqn,epsf]{article}
\def\Msun{\hbox{$M_\odot$}}

\def\deg{\hbox{$^\circ$}}

\title{The Emission from Post-Shock Flows in mCVs}

\author{
Mark Cropper\address{Mullard Space Science Laboratory, University
College London, \\ Holmbury St Mary, Dorking, Surrey RH5 6NT},
Kinwah Wu\address{Research Centre for Theoretical Astrophysics,
School of Physics, \\ Sydney University, Sydney, NSW 2006, Australia},
Gavin Ramsay$^{\rm a}$
}

\begin{document}
% typeset front matter
\maketitle

\begin{abstract}
We re-examine the vertical structure of the post-shock flow in the accretion
region of mCVs, and the X-ray emission as a function of
height. We then predict X-ray light curves and phase-resolved spectra, taking
into account the vertical structure, examine the implications and check whether
the predicted heights are compatible with observation.
\end{abstract}

\section{Introduction}

The accretion processes in magnetic CVs are considered to be important
astrophysically because they are one-dimensional and
quasi-collimated. This advantage is accompanied by relatively tractable
physical conditions in the accretion region coupled with a wide range of
diagnostic observations over a large range in wavelength, including polarised
optical emission.

Whether directly in a stream or from the inner edge of the accretion disk, the
accretion flow is constrained to follow the magnetic field lines to the surface
of the white dwarf primary. There material in supersonic free-fall is brought
to rest, forming a stand-off shock, with the temperature of order the
randomised free-fall velocity $\sim10-50$ keV. Behind this shock is a column of
hot plasma, which can settle onto the white dwarf only by cooling. In mCVs this
occurs by emitting bremsstrahlung radiation and, if the magnetic field is
sufficiently strong, by cyclotron radiation. 

The characteristics of the accretion column depend principally on the mass of
the white dwarf (which, via the free-fall velocity, sets the post-shock
temperature), on the local accretion rate (which affects the density, and
therefore the rate at which the region can radiate) and on the magnetic field
(which determines how much additional cooling will be produced due to cyclotron
radiation). With appropriate boundary conditions, the equations of conservation
of mass, momentum and energy can be used to calculate the temperature, density
and other hydrodynamic variables as a function of height within the column. How
easy it is to do this depends on the assumptions of the model. An analytic
solution (Aizu 1973) can be found in the case of bremsstrahlung radiation as
the only heating/cooling mechanism, with no gravity. Otherwise, at present the
equations have to be computed numerically (eg. Imamura \& Durisen 1983 and many
others), but for some cases the situation is better, with the existence of
closed integral forms (eg. Wu, Chanmugam \& Shaviv 1994). Models now include
the cooling by cyclotron emission, the effect of the variation of gravity
within the flow, the formulation of 2-fluid descriptions for ions and
electrons, and time dependent forms also exist.

With a prescription of temperature and density as a function of height within
the flow, the X-ray emission spectrum from this flow can be calculated. This is
typically done by the summation of optically thin emission spectra calculated
using standard optically thin models (eg. Cropper, Ramsay \& Wu 1998). By
comparing these multi-temperature model spectra to X-ray data, the basic
parameters of the model such as the white dwarf mass and the mass transfer rate
can be determined. Generally good fits are obtained, and a number of
cross-checks have been made with other methods of white dwarf mass
determination (Ramsay et al. 1998): these suggest that the method may yield
mass determinations which are higher than generally expected. Some difficulties
are clearly caused by inappropriate absorption models used in the fits to the
X-ray data, even at high energies (Done \& Magdziarz 1998). Other difficulties
may be caused by inappropriate assumptions within the modelling, which we now
explore.

\section{Emission at the Base of the Accretion Region}

One of the assumed boundary conditions in all the above work is a cold, hard
white dwarf surface. Because bremsstrahlung emission is proportional to the
square of the density, and towards the base of the region the density is rising
rapidly (to infinity at zero temperature) {\it
most} of the emission from the column is from its base (Fig. 1). 
For a 1.0\Msun\/ white dwarf accreting at 1 g/s/cm$^{2}$, $>50$\% of
the emission is in the lowest 5\% of the column at typical {\it Rosat} energies
of 1 keV. In $\nu$F$_{\nu}$ terms (where a direct comparison cam be made of the
energy output), most of the energy is emitted at $\sim 8$ keV -- see
Fig. 2. However, even at 8 keV, or in bolometric terms (Fig. 1), $\sim$ 1/3 of
the emission is from the bottom 5\%.

\begin{figure}
\setlength{\unitlength}{1cm}
\begin{picture}(8,7.0)
\put(-0.5,-0.93){\includegraphics{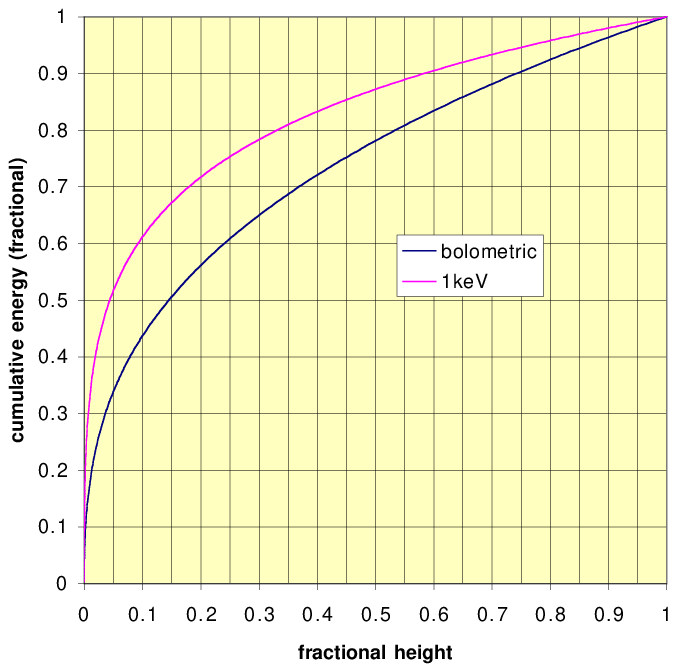}}
\put(8.3,-3.6){\includegraphics{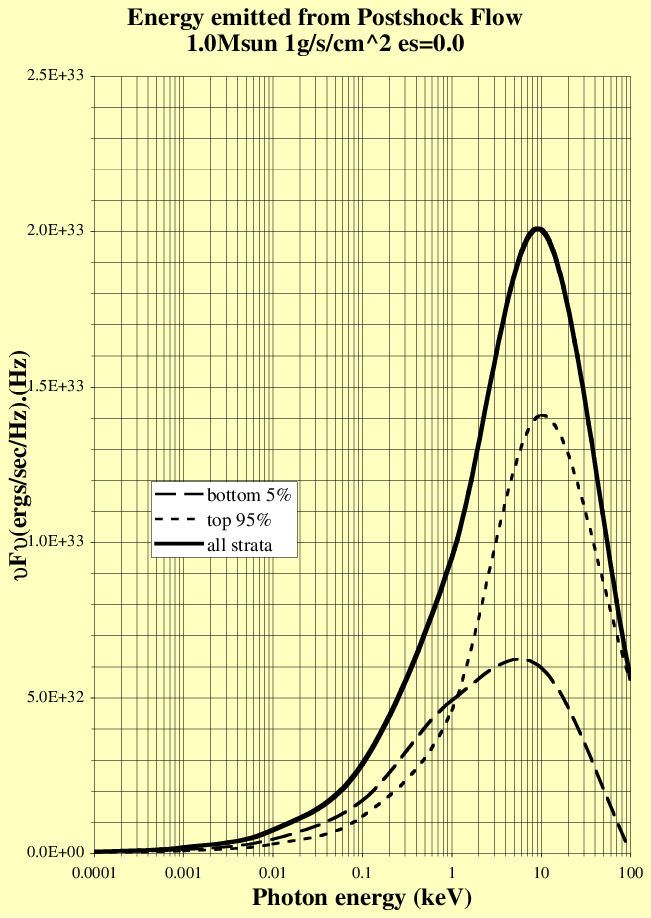}}
\end{picture}
\parbox{8.0cm}{\small {\bf Figure 1} (above) 
The cumulative energy fraction as a 
function of height in the accretion column at 1 keV and in total bolometric
luminosity, for a 1.0
\Msun\/ white dwarf with no cyclotron cooling accreting at 1 g/s/cm$^{2}$.\\ 
{\bf Figure 2} (right) The energy spectrum in $\nu$F$_{\nu}$ emitted from
the bottom 5\% of the column in Fig. 1, the top 95\% and in total.}
\vspace*{-6mm}
%\hspace*{1.5cm}
%\parbox{8.0cm}{{\bf Figure 2} blah, blah}
\end{figure}

It is worth appreciating the degree to which the emission is most intense at
the base of the region. This is evident in Fig. 3, which shows the spectrum
from 0.1 to 10 keV for the 1 \Msun\/ case above as a function of height plotted
on a linear greyscale. The high intensity of emission near the x-axis (the
surface of the white dwarf) is clear. What is even more striking from this
figure is that the emission from the lower part of the column exceeds that from
higher up even to 20 keV. Only at energies approaching the shock temperature
($\sim50$ keV for the 1.0 \Msun\/ white dwarf) does the emission from the top
of the column immediately below the shock approach that from the base near the
white dwarf.

\begin{figure}
\setlength{\unitlength}{1cm}
\begin{picture}(8,6.7)
\put(-1.8,-5.5){\includegraphics{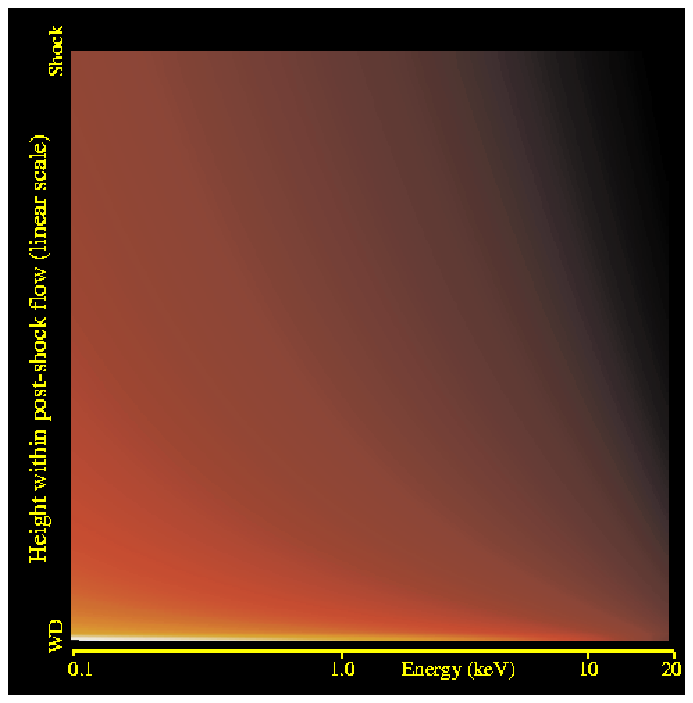}}
\put(6.7,-2.7){\includegraphics{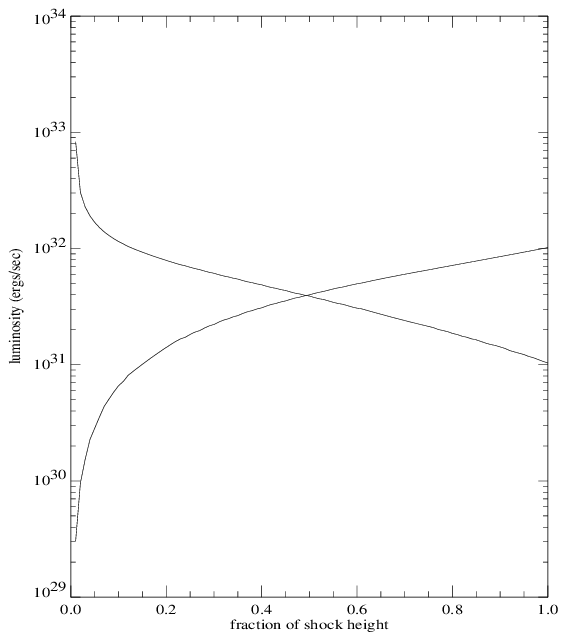}}
\end{picture}\\
\parbox[b]{7.5cm}{\small {\bf Figure 3} 
The spectrum from 0.1 to 20 keV plotted 
as a linear greyscale in intensity as a function of height within the column.
System parameters are a 1.0 \Msun\/ white dwarf accreting at 1 g/s/cm$^2$.}
\hspace*{0.8cm}
\parbox[b]{7.35cm}{\small {\bf Figure 4} 
The bremsstrahlung (increasing towards the
white dwarf surface) and cyclotron (decreasing) luminosity as a function of
height for the system in Fig. 3, but with a 23 MG field and a fractional area
over which accretion takes place of $f=0.008$.}
\vspace*{-6mm}
\end{figure}

The implications of this are twofold. Firstly there is the practical problem of
the sampling of the strata at the base of the column in the modelling routines:
those routines which sample the base of the column more finely will tend to
predict softer spectra, and therefore will predict higher mass white dwarfs in
order to match any observed spectrum, than those with coarser sampling. The
selection of the centre, top or bottom of any stratum to designate its height
will also have an effect, as the lower strata contribute so heavily to the
overall spectrum.

Secondly, and of more interest scientifically, the boundary condition used for
the base of the column is clearly inappropriate. At this point the hydrodynamic
flow ought to be matched to the hydrostatic atmosphere, and the conduction of
heat across this interface included in the calculations for the temperature and
density structure of the column (see Wu 1999). Inappropriate modelling of the
base of the column will lead to larger corrections to the emitted spectrum than
formerly appreciated, even in the 1 -- 10 keV range, in the direction of making
the spectrum harder and leading to lower mass white dwarfs from fits to data.

\section{Soft emission}

Towards the base of the column opacity effects will become important. 
Observing from infinity into the column, and 
ignoring the effects of any pre-shock flow, it can be shown that for a 1.0
\Msun\/ white dwarf accreting at 1 g/s/cm$^2$ over a fraction $f=0.008$ of its
surface, the free-free optical depth within the flow is negligible down to
$10^{-4}$ of the fractional height in both transverse and parallel
directions. However, electron scattering can be important (Rosen
1992). For the column above, electron scattering optical depths of unity in the
transverse direction are reached at a fractional height of 0.02. This is
significant when taking into consideration that the bulk of the emission occurs
from these heights. Although wavelength independent, the effect of the electron
scattering is to scatter photons from the lower strata out of the lateral line
of sight into other lines of sight, making the spectrum from the column as a
whole harder and fainter when seen side on and softer and brighter when seen at
other angles.

In addition to the continuum opacity effects, line and edge opacities will
become increasingly important in the base. Even without a detailed calculation,
it is possible to predict that from a side on view, for a cylindrical
column there will be a domed surface at the base where the optical depth is
unity. The spectrum of the photons from within this surface will be degraded to
a black body. Again the implications of this are twofold: firstly, these
photons are from the accretion flow, and in terms of the modelling of this
region belong to the soft end of the spectrum from this flow. Observers on the
other hand will assign these photons to the ``black body'' component
reprocessed from the heated photosphere. The effect of this is for the observer
to measure an excess of soft X-rays over what is expected from the standard
column (Lamb \& Masters 1979). At some level this effect must contribute to the
``soft X-ray problem''.  The extent to which this is the case will depend on
the details of the calculations for the opacities, but again, the effect will
be more important than formerly appreciated because most of the emission in the
range of imaging X-ray instruments is from the base of the column.

The second implication is that at soft energies, such as those seen by {\it
ROSAT} or {\it EUVE}, the accretion column will appear dome-shaped when seen
side on. Such an appearance has been suggested many times from {\it EUVE}
observations eg. Sirk \& Howell (1998). 

Finally, although limited work has been carried out in this area, the accretion
flow will produce a heated photosphere in the region of the base of the
accretion column (Litchfield \& King 1990) which will also modify the soft
X-ray emission from that region by absorption and scattering to a greater
degree than previously appreciated.

\section{Blobs}

Flares in X-ray and optical light curves are generally
considered to be the result of ``blobs'' 
or density enhancements in the flow. The
standard explanation for the soft X-ray problem is that it results from
energy deposited by blobs buried in the photosphere (Kuijpers \& Pringle 1982,
Frank, King \& Lasota, 1988).  It is clear that such blobs do exist, and the
arguments above do not eliminate their role in producing a soft
excess. However, they suggest that this excess can be produced without
requiring the majority of the blob to be buried -- only the base from where
most of the emission emanates.  This reduces the requirements on the density
spectrum of the blobs, and is consistent with the existence of optical flaring
from these systems (eg. VV Pup, Cropper \& Warner 1986). Because flares can
clearly be isolated to the column (from eclipse studies, Harrop-Allin et al.
1999) they are cyclotron radiation, which is emitted preferentially from the
hot upper parts of the column (Fig. 4). This region cannot be buried if optical
flares are observed.

Ramsay et al. (1994) and Beuermann \& Schwope (1994) found a correlation
between soft X-ray excess and magnetic field strength. The standard explanation
for this is that denser blobs are formed in stronger field systems, but the
complexity of the threading region prevents (at present) an analysis to
determine why or whether this may be the case. Alternatively we may explain it
in the framework of the standard model of the accretion column above: a blob
will cause a local density enhancement, which will reduce the column height;
the base of the region will then be relatively more buried than a neighbouring
region of greater height or the surrounding photosphere, so that more blackbody
soft X-rays are emitted. In addition, because of the higher local density,
the domed optical depth of unity surface encompasses more of the base of the
column than in neighbouring regions, resulting in more blackbody soft X-ray
flux.

Extrapolation of the argument to IPs where column height is expected to be tall
(because the local accretion rate is probably smaller than polars and there is
no cyclotron cooling) would suggest that this is one contributor to the absence
of a soft component in these systems, although local and interstellar
absorption is likely to play a greater role.

\section{Predictions from the height of the accretion column}

The next generation of X-ray observatories such as {\it XMM} and {\it Chandra}
will potentially provide phase resolved spectra of mCVs as the accretion column
rotates into view over the limb of the white dwarf. The rise in X-ray flux will
depend on both an increasing fraction of the height of the column becoming
visible, and on an increasing fraction of the area over which accretion is
occuring. We can easily generate the phase resolved spectra of an accretion
column accreting over any prescribed area onto the white dwarf by eliminating
the spectra from those strata which are out of view at any phase from the
spectral summation. Fig. 5 shows the prediction for a small accretion area,
using the models of Cropper et al. (1999), which include the modification
of the height due to the changing gravitational potential. The phase-resolved
spectra during rise to and fall from maximum can be compared to data to extract
the emission characteristics, and therefore the temperature and density as a
function of height. These fits can serve as a relatively direct check on the
validity of the models for the accretion column.

\begin{figure}[h]
\setlength{\unitlength}{1cm}
\begin{picture}(8,7.3)
\put(-0.5,-0.4){\includegraphics{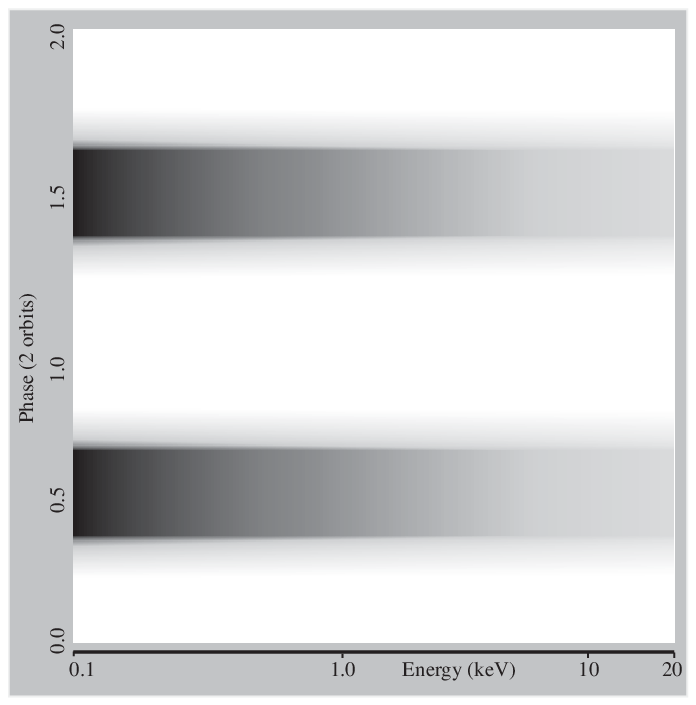}}
\put(6.4,-4.5){\includegraphics{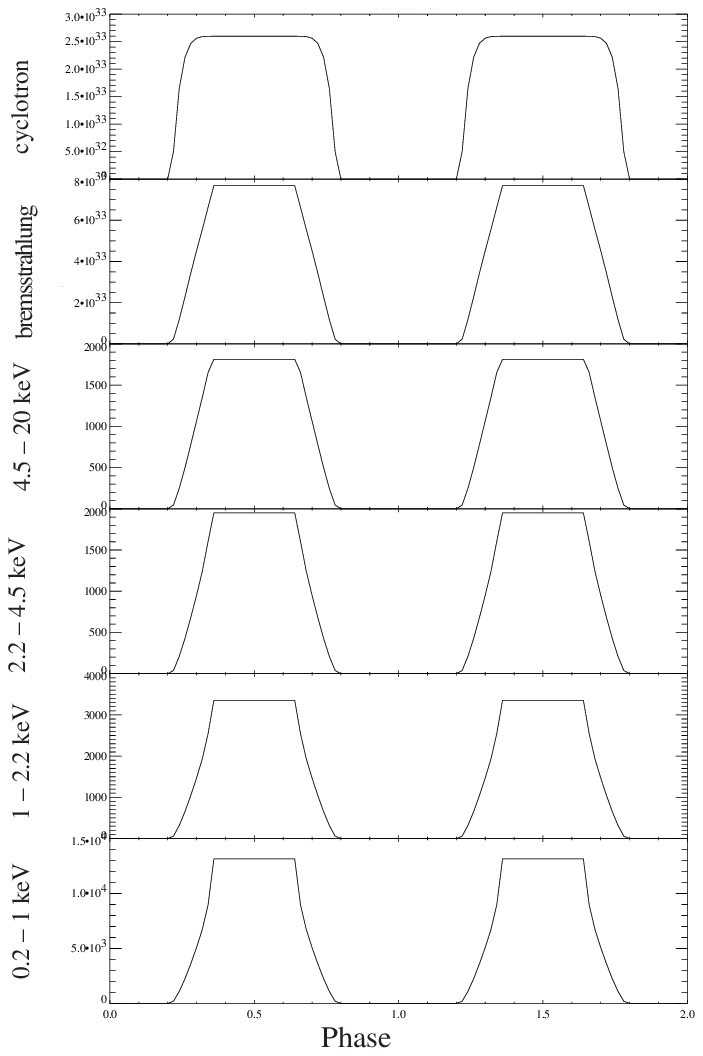}}
\end{picture}
%\end{center}
\parbox{8.0cm}{\small {\bf Figure 5} (above)
The phase-resolved spectra from 0.1 to 20
keV from the system in Figure 3, with a dipole offset of $\beta=134\deg$ viewed
at an inclination of $\iota=56\deg$. Phases are repeated vertically for
clarity.\\
{\bf Figure 6} (right)
Cyclotron, bolometric X-ray and X-ray light curves in the
4.5--20 keV, 2.2--4.5 keV, 1--2.2 keV and 0.2--1 keV (bottom) bands for the
system in Fig. 5. Phases are repeated for clarity.}
\vspace*{-6mm}
%\hspace*{1.5cm}
%\parbox{8.0cm}{{\bf Figure 2} blah, blah}
\end{figure}

X-ray light curves can be generated by summing flux from specified bandpasses
from Fig. 5. Because we include the cyclotron cooling in the models, it is also
possible to calculate the light curve of the total cyclotron emission. These
are shown for the particular inclination and dipole offset ($\iota=56\deg$,
$\beta=134\deg$, Cropper 1986) appropriate to ST LMi, a 2-pole polar (Fig. 6).
The duration of the bright phase is shorter for soft X-rays, which are emitted
at the base of the column, than that for cyclotron radiation, emitted near the
shock (Fig. 4), as this appears from behind the limb at earlier phases. 

These light curves can also be compared with X-ray data to extract information
on the vertical and spatial structure of the accretion region. Using the models
above of the vertical structure and X-ray emission as a function of mass
transfer rate per unit area, maps of the accretion rate can be generated using
maximum entropy techniques along the lines of those in polarised optical
emission (Stokes imaging, Potter et al. 1998), taking into account the
3-dimensional nature of the column. If Stokes imaging maps are available, these
can be compared.  Moreover, current Stokes imaging maps assume that the
cyclotron radiation is emitted on the surface of the white dwarf, whereas Fig.
4 indicates that emission is near the shock. This could lead to inaccurate
maps: polarisation reversals such as those at the end of the bright phase in
some systems could be the result of viewing a tall column from underneath, 
rather than or in addition to accretion
along non-radial fieldlines.

Finally, we can use the predictions of the accretion column models to
investigate whether accretion columns are ``tall and thin'' or
``pillbox-shaped''. This issue from the early epochs of mCV research has
appeared to be settled by observation in terms of the latter, particularly
based on observations of ST LMi, where no polarisation reversals are seen at
the beginning and end of the bright phase (eg. Cropper 1986). The predictions
for this geometry in Figs. 5 and 6 indicate that the column is sufficienty tall
to be seen from ``underneath'' for a considerable range of phase during the
rise to and fall from maximum light. These, however, are for a 1.0 \Msun\/
white dwarf and no cyclotron cooling. When an appropriate mass (0.45 \Msun\/,
Mukai \& Charles 1987) and magnetic field (12 MG, Ferrario, Bailey \&
Wickramasinghe, 1993) are used, the height of the column drops suffiently that
the duration of the rise phase is small enough not to be visible at the phase
resolution of the polarisation data. This compatibility indicates that the
height of the column in this system is indeed low from both theoretical
prediction and observations. However, in those systems with higher mass white
dwarfs and/or lower magnetic fields the column will, contrarily, be tall with
implications for interpretation of both X-ray and optical data as discussed
above.

\end{document}